\documentclass[letterpaper,twocolumn,showpacs,pra,aps]{revtex4-1}

\usepackage{amsmath}  
\usepackage{amsfonts} 
\usepackage{graphicx} 
\usepackage{amssymb}
\usepackage{subfig}
\usepackage{float}
\usepackage{xcolor}

\begin{document}

\title{Splashes in isotropic media}

\author{Eugene B. Kolomeisky}

\affiliation{Department of Physics, University of Virginia, P. O. Box 400714,
Charlottesville, Virginia 22904-4714, USA}

\date{\today}

\begin{abstract}
The response of a weakly absorbing isotropic medium to a sudden localized perturbation (a "splash") is explained within the framework of linear response theory.  In this theory splashes result from the interference of the collective excitations of the medium, with the outcome determined by the interplay between their phase and group velocities as well as the sign of the latter. The salient features of splashes are controlled by the existence of extremal values of the phase and the group velocities: the group velocity gives the expansion rate of the locus of the points where new wavefronts nucleate or existing ones disappear, while the phase velocity determines the large-time expansion rate of a group of wavefronts.  If the group velocity is negative in a spectral range and takes on a minimal value within it, then \textit{converging} wavefronts will be present in the splash.  These results are relevant to the studies of several experimentally viable setups, such as a splash on the surface of deep water due to a small pebble or a raindrop, a splash in the two-dimensional electron gas caused by a short voltage pulse applied with the tip of a scanning tunneling microscope, or a bulk splash in superfluid $^{4}He$ due to formation of an electron bubble. Specifically, the gross features of a splash in superfluid $^{4}He$ are determined by \textit{five} extremal velocities.  Additionally, due to the existence of a negative group velocity spectral range, some of the wavefronts in the superfluid splash are converging.  
 
\end{abstract}


\maketitle

\section{Introduction}  

Recent years have seen major advances in imaging of the fluid density $n(\textbf{r},t)$ and velocity $\textbf{v}(\textbf{r},t)$ fields in neutral quantum liquids such as superfluid $^{4}He$ \cite{He41,He42,He43,He44}, superfluid $^{3}He$ \cite{He3}, atomic gas superfluids \cite{vapor}, and charged quantum liquids such as electrons in graphene \cite{imaging} and Cooper-pair liquids in superconductors \cite{super}.   These emerging capabilities open a door to visualization of a variety of effects, some of which have already been described \cite{Levitov,Kelvin_Mach}.  A theoretical study of a family of such effects, splashes, is given in this paper.

A splash is the response of the medium to a local perturbation of a short duration;  it is described by an initial value problem.  A familiar example from classical physics is the expanding pattern of annular waves caused by a small rock impacting a surface of calm water \cite{Lamb,Crapper,Stoker}.  Analogs of this phenomenon exist in quantum liquids.  For instance, a low-energy electron injected inside liquid $^{4}He$ triggers quick formation of a bubble around it \cite{Ferrell};  the reaction of the superfluid to the formation of the bubble is an example of a three-dimensional splash in a neutral quantum liquid.  The surfaces of superconductors and two-dimensional electron systems allow for a large degree of control over the perturbation.  Specifically, applying a short voltage pulse with the tip of a scanning tunneling microscope to a graphene sheet \cite{Eva_Andrei} can initiate a splash in the two-dimensional sea of Dirac electrons.

For any localized disturbance, the response of the medium is small at long times $t$ after the perturbation ceased to operate, and then is described by a linear theory.  In such a theory, developed below, splashes are a result of the interference of collective excitations of the medium;  the outcome is determined by their frequency spectrum $\Omega(\textbf{k})$ (here $\textbf{k}$ is the wave vector).  

Previous treatments of splashes focused on the water surface as a medium.  Here the relevant excitations are the capillary gravity waves whose dispersion law in the deep water (and incompressible fluid) limit has the form \cite{LL6}
\begin{equation}
\label{capillary_gravity}
\Omega^{2}(\textbf{k})=gk+\frac{\gamma}{\rho}k^{3}
\end{equation}  
where $g$ is the free-fall acceleration, $k=|\textbf{k}|$ is the wave number, $\gamma$ is the coefficient of surface tension of water and $\rho$ is the density of water.  
 
For $\gamma=0$ the initial value problem has been fully solved by Cauchy and Poisson (CP) \cite{Lamb}.  Specifically, the position of the $l$-th wavefront in the splash $r_{l}$ as measured from the point of impact in the large time limit $gt^{2}/r_{l}\gg1$ is given by the expression 
\begin{equation}
\label{CP_splash}
r_{l}=\frac{gt^{2}}{8\pi l}
\end{equation}
whose hallmark is accelerated expansion of the rings.

The parameters of the spectrum (\ref{capillary_gravity}) can be combined to form a spatial scale $\lambda$, the capillary length, and a time scale $\tau$, such as
\begin{equation}
\label{water_splash_scales}
\lambda=\left (\frac{\gamma}{g\rho}\right )^{1/2}=0.28~\text{cm},~~~\tau=\left (\frac{\gamma}{g^{3}\rho}\right )^{1/4}=0.017~\text{s}
\end{equation}
where the numerical values are for water at $20^{\circ}$C \cite{LL6}. The CP theory is only valid for wavelengths long compared to the capillary length.   When the scales (\ref{water_splash_scales}) are used as units of length and time (see below), the dispersion law (\ref{capillary_gravity}) acquires the parameter-free form
\begin{equation}
\label{cg_parameter-free}
\Omega^{2}=k+k^{3}   
\end{equation}
which means that there is more to the water splash than implied by the CP result (\ref{CP_splash}).

Kelvin \cite{Kelvin} pioneered a general method for analyzing splashes due to excitations with arbitrary dispersion law, and found that the dynamics of splashes is determined by the interplay between the phase and the group velocities.  The CP result (\ref{CP_splash}) has its simple form because the group velocity for gravity waves ($\gamma=0$ in Eq.(\ref{capillary_gravity})) is half the phase velocity.  

\begin{figure}
\includegraphics[width=1.0\columnwidth, keepaspectratio]{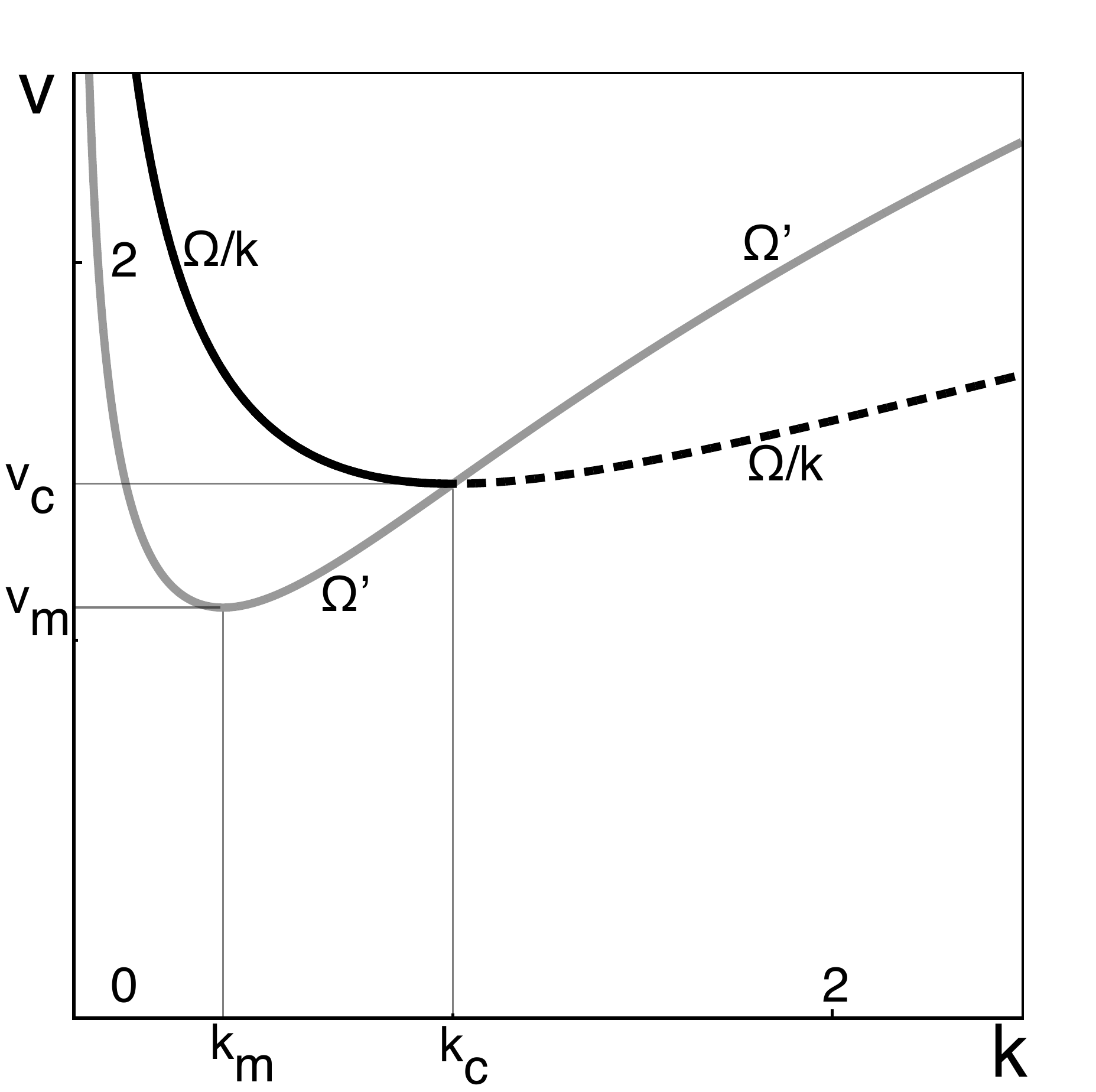} 
\caption{Group $\Omega'(k)$ and phase $\Omega(k)/k$ velocities versus the wave number $k$ in units of $l/\tau$ and $1/l$, respectively, Eq.(\ref{water_splash_scales}), for capillary gravity waves (\ref{cg_parameter-free}).} 
\label{phasegroup}
\end{figure}

For the general dispersion relation, Kelvin argued that interesting phenomena are sure to occur whenever there are extrema of the phase velocity, because then the phase velocity $\Omega/k$ and group velocity $d\Omega/dk\equiv \Omega'(k)$ are equal (indeed the condition $(\Omega/k)'=0$ is equivalent to $\Omega/k=\Omega'(k)$).  For the particular case of capillary gravity waves (shown in Figure \ref{phasegroup}), there is a minimum when $k=k_{c}=1$ and 
\begin{equation}
\label{Kelvin_velocity}
v=v_{c}=\sqrt{2}~~(23~\text{cm/s, physical~units}).
\end{equation}

The group velocity $\Omega'$ also has a minimum (Figure \ref{phasegroup}) at $k_{m}=(2/\sqrt{3}-1)^{1/2}\approx0.39$ corresponding to the velocity 
\begin{equation}
\label{minimal_velocity}
v_{m}=\sqrt{3}\left (\frac{2}{\sqrt{3}}-1\right )^{1/4}\approx1.08~ (18~\text{cm/s, physical~units})
\end{equation}
Observations show \cite{Crapper,Stoker} that a few seconds following the perturbation of a water surface, a quiescent region inside the annular waves is formed.  Rayleigh demonstrated \cite{Rayleigh} that this region expands with the constant velocity $v_{m}$  (\ref{minimal_velocity}).   Outside the quiescent region there are two systems of waves of different wavelength present at the same place.  In practice only one system is  visible, and Rayleigh conjectured that the other (corresponding to short waves of predominantly capillary origin) is rapidly damped.   
Rayleigh's conjecture has been justified by Lighthill \cite{Lighthill} who also observed that new wavefronts nucleate "from nowhere" at the boundary of the expanding region of calmed water.

Le M\'ehaut\'e \cite{Le} additionally argued that the waves in the annular region have a narrow range of wave numbers centered around $k_{m}$, Figure \ref{phasegroup}, corresponding to the minimum group velocity $v_{m}$ (\ref{minimal_velocity}).

Below we give a general theory of dynamics of wavefronts in splashes in the weakly absorbing isotropic medium and apply it to various cases.   While elaborating on Rayleigh's results \cite{Rayleigh} regarding the water spash, we expand on Lighthill's observation \cite{Lighthill}, showing that new wavefronts arise at the inner boundary of the annular region in pairs at equal time intervals.  We also support Kelvin's intuition regarding the significance of the minimum phase velocity $v_{c}$ (\ref{Kelvin_velocity}):  it sets the velocity of expansion of a group of capillary gravity rings in the long-time limit where the conjecture of Ref. \cite{Le} fails.   

This theory also applies to a plasmonic splash in a two-dimensional Fermi sea.  Here the relevant dispersion law is  \cite{Fetter,AFS,Hwang_D_Sarma,DasHwang,EBKJPS}:
\begin{equation}
\label{plasma}
\Omega^{2}(\textbf{k})=gk+u^{2}k^{2}
\end{equation} 
where $g$ (no longer the free-fall acceleration) and the speed of sound $u$ are determined by the equation of state of the electron gas \cite{EBKJPS}.  While the spectra (\ref{capillary_gravity}) and (\ref{plasma}) are the same in the long-wavelength limit, the remaining difference - $k^{3}$ versus $k^{2}$ contributions - makes the plasmonic splash a simpler version of its water counterpart as discussed below.  

A final application of the theory involves a splash in bulk superfluid $^{4}He$ whose elementary excitations exhibit a spectral region with negative group velocity \cite{LL9}.  A recent analysis of the wake patterns in this system \cite{GJJ} established that these excitations are responsible for features similar to the Kelvin ship wake.    Below it will be shown that negative group velocity excitations are responsible for \textit{converging} wavefronts in superfluid $^{4}He$ splashes.          

\section{Formalism}  

Regardless of its particular manifestation, splashes are described by linear response theory \cite{LL5,LL9,Pines_Nozieres}.  Let us suppose that every particle of the medium is perturbed by an external field of the potential energy $U(\textbf{r}, t)$.  Then the operator of the perturbation acting on the whole medium is 
\begin{equation}
\label{perturbation}
\hat{V}(t)=\int \hat{n}(\textbf{r},t)U(\textbf{r},t)d^{d}x
\end{equation}
where $\hat{n}(\textbf{r},t)$ is the Heisenberg density operator and $d$ is the space dimensionality (in classical linear water wave theory $n(\textbf{r},t)$ is the height of the water surface while $U(\textbf{r},t)$ is the excess pressure \cite{JCEBK}).  The Fourier transform of the induced density due to the perturbation is given by $\delta{n}(\omega,\textbf{k})=-\alpha(\omega,\textbf{k})U(\omega,\textbf{k})$ where $\alpha(\omega,\textbf{k})$ is a generalized susceptibility \cite{LL9,LL5, Pines_Nozieres} and $U(\omega,\textbf{k})$ is the Fourier transform of $U(\textbf{r},t)$.  Inverting the Fourier transform and specifying to the case of a point instantaneous source, $U(\omega,\textbf{k})=const$, the induced density will be given by
\begin{equation}
\label{splash}
\delta{n}(\textbf{r},t)\propto \int\frac{d\omega d^{d}k}{(2\pi)^{d+1}}\alpha(\omega,\textbf{k})e^{i(\textbf{k}\cdot\textbf{r}-\omega t)}.
\end{equation}
The dynamics of the wavefronts in the splash can be determined by using Kelvin's stationary phase argument \cite{Kelvin,Lamb}.  At positions and times such as the phase $f=\mathbf{k}\cdot\mathbf{r}-\omega t$ is large in magnitude, the exponential in (\ref{splash}) is highly oscillatory, and contributions of elementary plane waves interfere destructively leaving almost no net result, unless $\omega=\pm \Omega(\textbf{k})$ (which are the poles of the susceptibility $\alpha(\omega,\textbf{k})$  \cite{LL9,Pines_Nozieres}) and have a phase which is stationary with respect to $\mathbf{k}$.  Subjecting the phase
\begin{equation}
\label{splash_phase}
f=\textbf{k}\cdot\textbf{r}\mp\Omega(\textbf{k})t
\end{equation}
to the condition of stationarity $\nabla_{\textbf{k}}f=0$ one finds
\begin{equation}
\label{splash_stationary_phase}
\textbf{r}=\pm\nabla_{\textbf{k}}\Omega \cdot t.
\end{equation}
Since the phase $f$ is constant along the wavefronts, the last two equations can be solved to determine the wavefront dynamics in a parametric form.  In an isotropic medium they become 
\begin{equation}
\label{splash_parametric_radial}
r(k)=\frac{\Omega'}{\Omega' k-\Omega}f, ~~~~~~t(k)=\pm\frac{1}{\Omega' k-\Omega}f
\end{equation}
where the lower sign in the expression for $t(k)$ accounts for the possibility of a negative group velocity. 

Since $r$ and $t$ are positive, the phase $f$ is determined by the interplay between the phase and the group velocities, as well as by the sign of the latter.  Specifically, there are three possibilities:    
\begin{equation}
\label{phases}
  f = \left \{
\begin{aligned}
    &2\pi l, && \text{if}\ \Omega'>\Omega/k \\
    &-2\pi l, && \text{if}\ \ 0<\Omega'<\Omega/k\\
    &2\pi l,  && \text{if}\ \ \Omega'<0  
    \end{aligned} \right.
\end{equation} 
where $l$ is a positive integer.  The dynamics of the wavefronts of the last type is given by Eqs.(\ref{splash_parametric_radial}) with the lower sign chosen in the expression for $t(k)$;  otherwise, the expression for $t(k)$ with the upper sign should be used.

Several conclusions anticipating the gross features of splashes can be deduced from Eqs.(\ref{splash_parametric_radial}):  

(i)  When the equation $\Omega''=0$ has real solutions, i. e. the group velocity has an extremum $v=v_{m}$, the expressions for $r(k)$ and $t(k)$ have simultaneous extrema.  Then the equation $r=|v_{m}|t$ gives the locus of the points where new wavefronts nucleate or existing ones disappear.  When this happens at a nonzero $k=k_{m}$ which is not an end point of the spectrum, the wavefronts appear or disappear in pairs.  Since positions of extrema of $t(k)$ are $l$-independent, the wavefronts appear (or disappear) at regular time intervals
\begin{equation}
\label{interval}
t(k_{m})\equiv t_{m}=\frac{2\pi}{|v_{m} k_{m}-\Omega(k_{m})|}.
\end{equation}

(ii)  In the vicinity of an extremum of the group velocity, the spectrum can be approximated by its Taylor expansion
\begin{equation}
\label{Taylor}
\Omega(k)=\Omega(k_{m})+v_{m}(k-k_{m})+\frac{\Omega^{(3)}(k_{m})}{3!}(k-k_{m})^{3}
\end{equation}
Combining it with Eqs.(\ref{splash_parametric_radial}) and (\ref{phases}), one can then see that when the group velocity has a minimum ($\Omega^{(3)}(k_{m})>0$), that is negative, $v_{m}<0$, then $t(k)$ has a \textit{minimum} while $r(k)$ has a \textit{maximum} at $k=k_{m}$.  The consequence is that pairs of wavefronts nucleating with period $t_{m}$ (\ref{interval}) will be \textit{converging} toward the center of the splash.    

(iii)  The large $t$ limit is controlled by the points of the spectrum where the phase and the group velocities are equal;  these are also extrema of the phase velocity.  If this happens at $k=k_{c}$ with finite common velocity $v=v_{c}$, equation $r=v_{c}t$ gives the asymptotic large $t$ behavior of wavefronts whose wave numbers are close to $k=k_{c}$.  If $k_{c}=0$, then $v_{c}$ is the speed of sound $u$.  If $k_{c}=0$ and $v_{c}=0$ ($v_{c}=\infty$), then the asymptotic large $t$ expansion of wavefronts is sub-ballistic (super-ballistic). 

(iv)  Eqs.(\ref{splash_parametric_radial}) and (\ref{phases}) imply that if $r/l$ and $t/l$ are used as variables to represent splashes, visual complexity of their original $r(t)$ patterns is reduced since all the wavefronts of given family (according to Eq.(\ref{phases})) "collapse" onto a single curve (or a pair of curves) representing that family. 

\section{Applications}

 We now proceed to selected applications of the general results (\ref{splash_parametric_radial}), (\ref{phases}) and (\ref{interval}).  

\subsection{Acoustic spectrum}

When the excitation spectrum is linear in the wavenumber $k$,
\begin{equation}
\label{sound}
\Omega=uk,
\end{equation}
the phase $\Omega/k$ and the group $\Omega'$ velocities are equal to $u$ for all $k$.  According to  Eqs.(\ref{splash_parametric_radial}) this is the marginal case.  The well-known outcome $r=ut$ then follows from Eq.(\ref{splash_stationary_phase}):  there is only one wavefront propagating away from the point of disturbance with the speed of sound.

\subsection{Gravity waves}  

For gravity waves ($\gamma=0$ in Eq.(\ref{capillary_gravity})) the group velocity is always smaller than the phase velocity, thus implying, Eq.(\ref{phases}), that $f=-2\pi l$. Then according to Eq.(\ref{splash_parametric_radial}) one finds that $r=2\pi l/k$ and $t=4\pi l/\sqrt{gk}$; combining them recovers the CP result (\ref{CP_splash}).  This is an example of a super-ballistic expansion.  

\subsection{Capillary waves}  

For capillary waves ($g=0$ in Eq.(\ref{capillary_gravity})) the group velocity is always larger than the phase velocity, thus implying, Eq.(\ref{phases}), that $f=2\pi l$. Then according to Eq.(\ref{splash_parametric_radial}) one finds that $r=6\pi l/k$ and $t=4\pi l/\sqrt{\gamma k^{3}/\rho}$.  Eliminating the wave number $k$ one obtains
\begin{equation}
\label{capillary_splash}
r_{l}=3\left ( \frac{\pi \gamma lt^{2}}{2\rho}\right )^{1/3},
\end{equation}
i.e. the expansion is sub-ballistic, $r_{l}\propto t^{2/3}$.  

\subsection{Capillary gravity waves}  

\begin{figure}
\includegraphics[width=1.0\columnwidth, keepaspectratio]{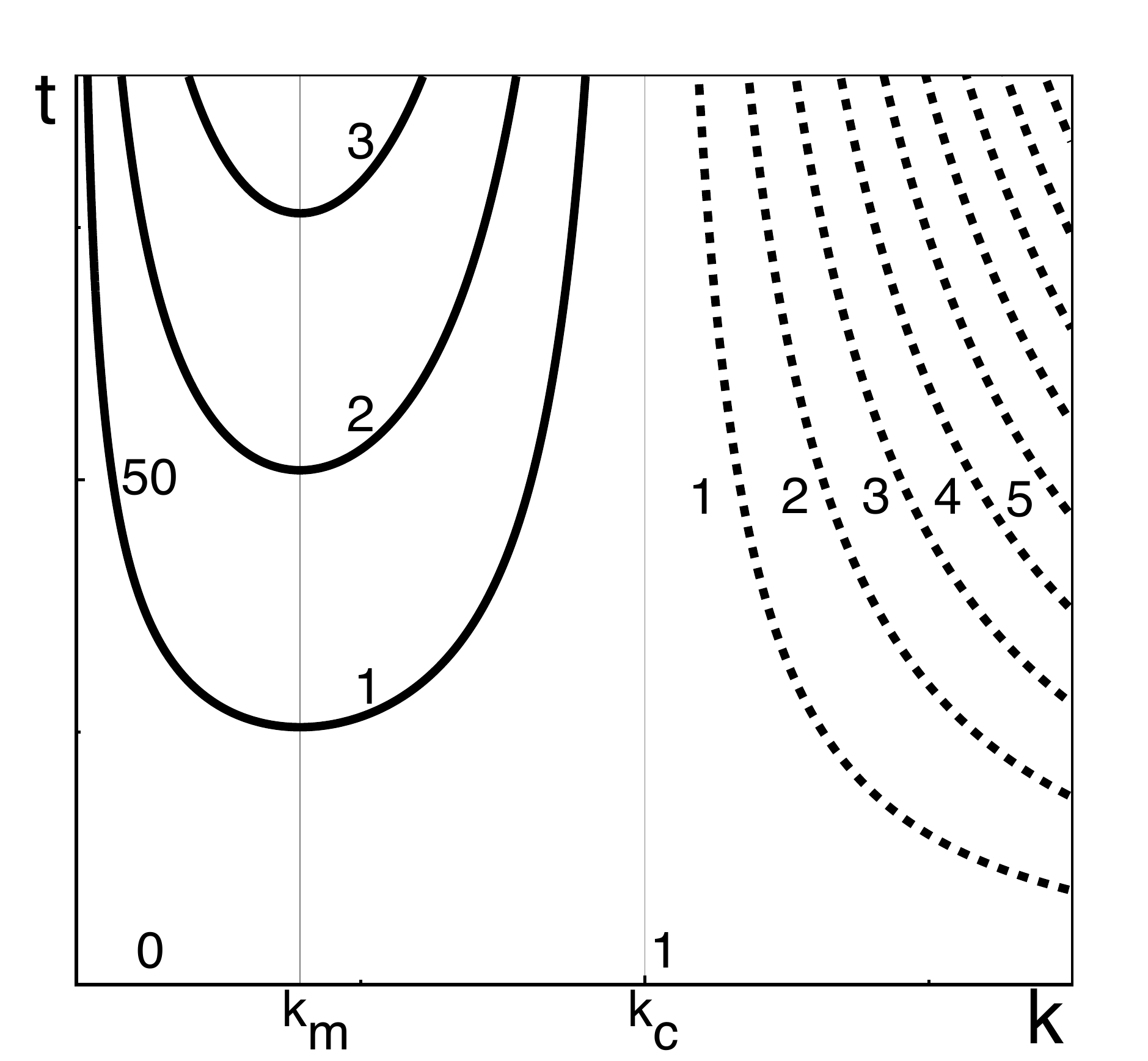} 
\caption{Dependences of $t(k)$, Eqs.(\ref{splash_parametric_radial}) and (\ref{phases}), for the capillary gravity waves  (\ref{cg_parameter-free}).  The legend and line styling are the same as in Figure \ref{phasegroup}.  Numbers next to the curves are values of integer $l$.} 
\label{tofk}
\end{figure}

\begin{figure}
\includegraphics[width=1.0\columnwidth, keepaspectratio]{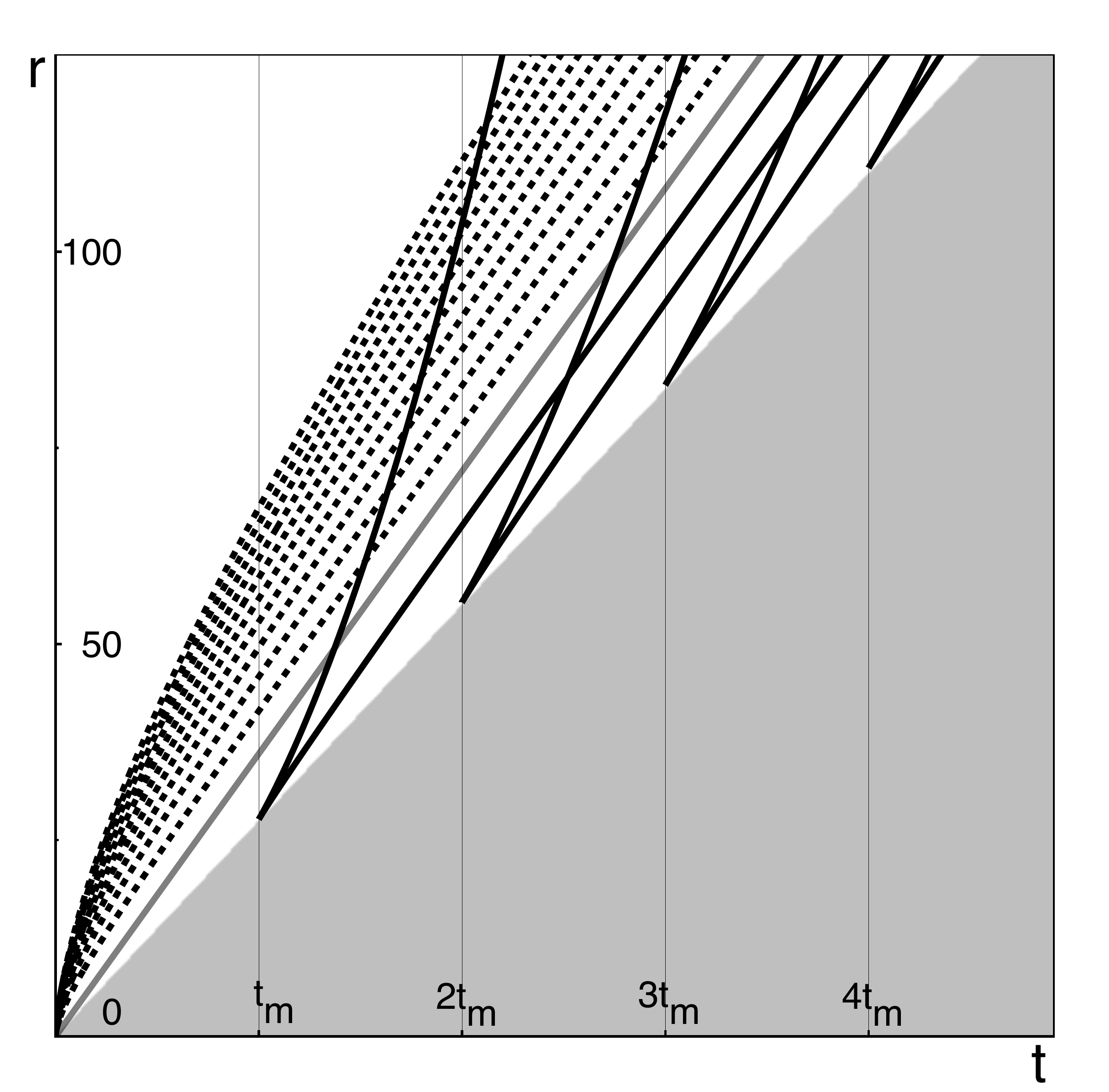} 
\caption{Radii of annular capillary gravity wavefronts vs time following a sudden localized perturbation of the water surface, in units of Eq.(\ref{water_splash_scales}) for several values of $l$ according to Eqs.(\ref{cg_parameter-free}), (\ref{splash_parametric_radial}), (\ref{phases}) and (\ref{water_periodicity}).  The line styling is coordinated with Figures \ref{phasegroup} and \ref{tofk}.  The greyscale line $r=v_{c}t$ separates annular rings made by short $k>k_{c}$ (dashed) and long $k<k_{c}$ waves.  The shaded grey region of calmed water expands with the velocity $v_{g}$ (\ref{Kelvin_velocity}).} 
\label{wsplash}
\end{figure}

With capillarity included (\ref{cg_parameter-free}), the first two possibilities of Eq.(\ref{phases}) are realized.  To understand the dynamics of the wavefronts, in Figure \ref{tofk} we plotted corresponding $t(k)$ dependences (\ref{splash_parametric_radial}) for several values of $l$ (the $r(k)$ dependences are not shown; they are qualitatively the same).
We observe that at times smaller than 
\begin{equation}
\label{water_periodicity}
t_{m}=2^{3/2}\pi\frac{(1+2/\sqrt{3})^{1/4}}{1-1/\sqrt{3}}\approx 25.47 (0.43~\text{s, physical~units})
\end{equation}
given by Eq.(\ref{interval}), the equation for $t(k)$ (\ref{splash_parametric_radial}) only has solutions corresponding to short waves $k^{(l)}(t)>k_{c}$ where $f=2\pi l$.  In the $t\ll t_{m}$ limit the effects of gravity are negligible, and evolution of these annular rings follows Eq.(\ref{capillary_splash}).  As $t\rightarrow \infty$ the solutions $k^{(l)}(t)$ asymptotically approach $k=k_{c}$ from above.  The corresponding rings expand with the velocity $v_{c}$ (\ref{Kelvin_velocity}) corresponding to zero of the denominator in Eqs.(\ref{splash_parametric_radial}).  The behavior for arbitrary $t$ is displayed in Figure \ref{wsplash} by a series of dashed wavefronts;  line styling is coordinated with the $k>k_{c}$ regions in Figures \ref{phasegroup} and \ref{tofk}.  As in the purely capillary case (\ref{capillary_splash}), the number of these wavefronts is infinite and they extend over the whole surface; this is an artifact of the incompressible liquid approximation.  

At $t=t_{m}$ (\ref{water_periodicity}) the equation for $t(k) $ (\ref{splash_parametric_radial}) acquires an additional solution $k=k_{m}$ which for $t>t_{m}$ bifurcates into two:  $k_{-}(t)<k_{m}$ and $k_{+}(t)>k_{m}$.  As the time progresses, the first of these tends to zero, $k_{-}(t\rightarrow \infty)\rightarrow 0$ where Eqs.(\ref{splash_parametric_radial}) reduces to the CP result (\ref{CP_splash}) for $r_{1}$.  On the other hand, as $t\rightarrow \infty$ the second solution $k_{+}(t)$ asymptotically approaches $k=k_{c}$ from bellow;  corresponding ring expands with the constant velocity $v_{c}$ (\ref{Kelvin_velocity}).  For arbitrary $t\geqslant t_{m}$ this is shown in Figure \ref{wsplash}:  "nucleation" at $t=t_{m}$ followed by bifurcation into two branches.

At $t=2t_{m}$ (\ref{water_periodicity}) the equation for $t(k)$ (\ref{splash_parametric_radial}) acquires yet another solution $k=k_{m}$ that for $t>2t_{m}$ bifurcates into two.  Their evolution repeats what was already found for the first pair of solutions $k_{-,+}(t)$.  Generally, new wavefronts are created periodically in $0.43~\text{s}$ intervals (\ref{water_periodicity}) followed by bifurcation into two, one of which, at $t$ large, approaches the CP result (\ref{CP_splash}) while the other expands with the velocity $v_{c}=23~cm/s$ (\ref{Kelvin_velocity}).  These wavefronts given by $f=-2\pi l$ solutions to Eqs.(\ref{splash_parametric_radial}), are shown in Figure \ref{wsplash}.  The shaded grey region of calmed water expands with the constant velocity $v_{g}=18~cm/s$ (\ref{Kelvin_velocity}).  In practice it should become clearly defined in a time interval estimated as several $t_{m}$ (\ref{water_periodicity}), i.e. several seconds, which explains observations.

The greyscale line $r=v_{c}t$ is shown for reference; it separates the two $f=\pm 2\pi l$ regimes discussed earlier.

Despite their ubiquity, systematic quantitative studies of splashes on deep water in the linear regime have been lacking.  Possible reasons for this have already been given by Rayleigh \cite{Rayleigh}: the short waves of capillary-gravity origin represented in Figure \ref{wsplash} by the "dashed" wavefronts may be rapidly damped, and the length scale for full development of the splash pattern may be inconveniently large.

Both of these obstacles can be overcome if instead of water one uses superfluid $^{4}He$.  Damping is eliminated in the superfluid state, and the coefficient of surface tension of the superfluid extrapolated to zero temperature, $\gamma=0.37~erg/cm^{2}$ \cite{Atkins}, is about 200 times smaller than that of water.  Given the density of the superfluid $\rho=0.145~g/cm^{3}$ \cite{LL9},  the $^{4}He$ counterparts of the capillary length and the time scale (\ref{water_splash_scales}) can be found as
\begin{equation}
\label{He_scales}
\lambda^{(He)}=0.051~\text{cm},~~~\tau^{(He)}=0.0072~\text{s}.
\end{equation}
Since the capillary length of water (\ref{water_splash_scales}) is five times larger, many more wavefronts will be present within the same observation area in the case of the superfluid.  Moreover, the velocity unit in the case of superfluid $^{4}He$, $\lambda^{(He)}/\tau^{(He)}=7~cm/s$, is about a half of that of water $\lambda/\tau=16~cm/s$ (\ref{water_splash_scales}).  

\subsection{Plasma waves in a two-dimensional electron gas} 

\begin{figure}
\includegraphics[width=1.0\columnwidth, keepaspectratio]{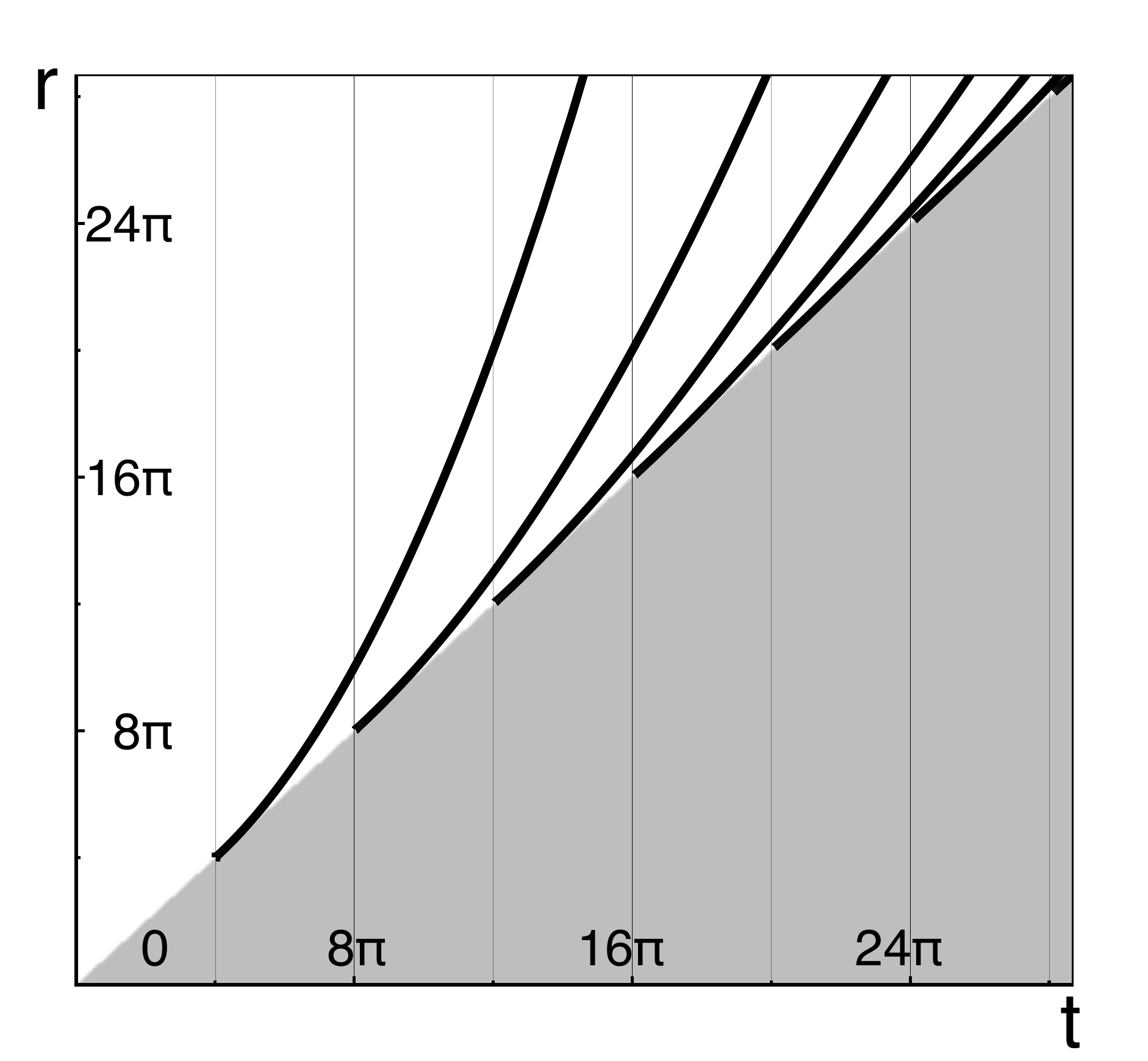} 
\caption{(Color online) Radii of the annular plasma wavefronts vs time (in units of $u^{2}/g$, the screening length, and $u/g$, respectively) following a sudden localized perturbation of a two-dimensional electron gas, as described by  Eqs.(\ref{splash_parametric_radial}), (\ref{phases}) and (\ref{plasma}).  The region of calmed electron liquid $r<ut$ is shaded grey.} 
\label{esplash}
\end{figure}

Evaluation of Eq. (\ref{plasma}) shows that both the group $\Omega'$ and the phase $\Omega/k$ velocities are monotonically decreasing functions of $k$ which asymptotically approach the speed of sound as $k\rightarrow \infty$, leading to the results that $v_{m}=u$ and $k_{m}=\infty$.  Since $\Omega'<\Omega/k$, Eq.(\ref{phases}) further implies that $f=-2\pi l$.  The second of Eqs.(\ref{splash_parametric_radial}) then becomes $t=4\pi l (u/g) \sqrt{1+g/u^{2}k}$.  The consequence is that Eqs.(\ref{splash_parametric_radial}) acquire solutions only for $t\geqslant t_{m}=4\pi u/g$.  The first of these is a monotonically increasing function of time shown in Figure \ref{esplash}
as the leftmost curve approaches the CP result (\ref{CP_splash}) for $r_{1}$ for $t\rightarrow \infty$.

More generally, new wavefronts are created periodically at times $t=lt_{m}=4\pi lu/g$.  Their evolution is shown in Figure \ref{esplash};  as $t\rightarrow \infty$ they approach the CP result (\ref{CP_splash}).  These wavefronts are found at space-time locations $r\geqslant ut$.  The region of calmed electron liquid $r<ut$ is shaded grey.  

The annular waves in Figure \ref{esplash} are counterparts to the accelerating wavefronts found in the water splash, Figure \ref{wsplash}.  The central difference is that annular waves in the two-dimensional electron gas are created one at a time. 

\subsection{Elementary excitations in superfluid $^{4}He$}  

\begin{figure}
\includegraphics[width=1.0\columnwidth, keepaspectratio]{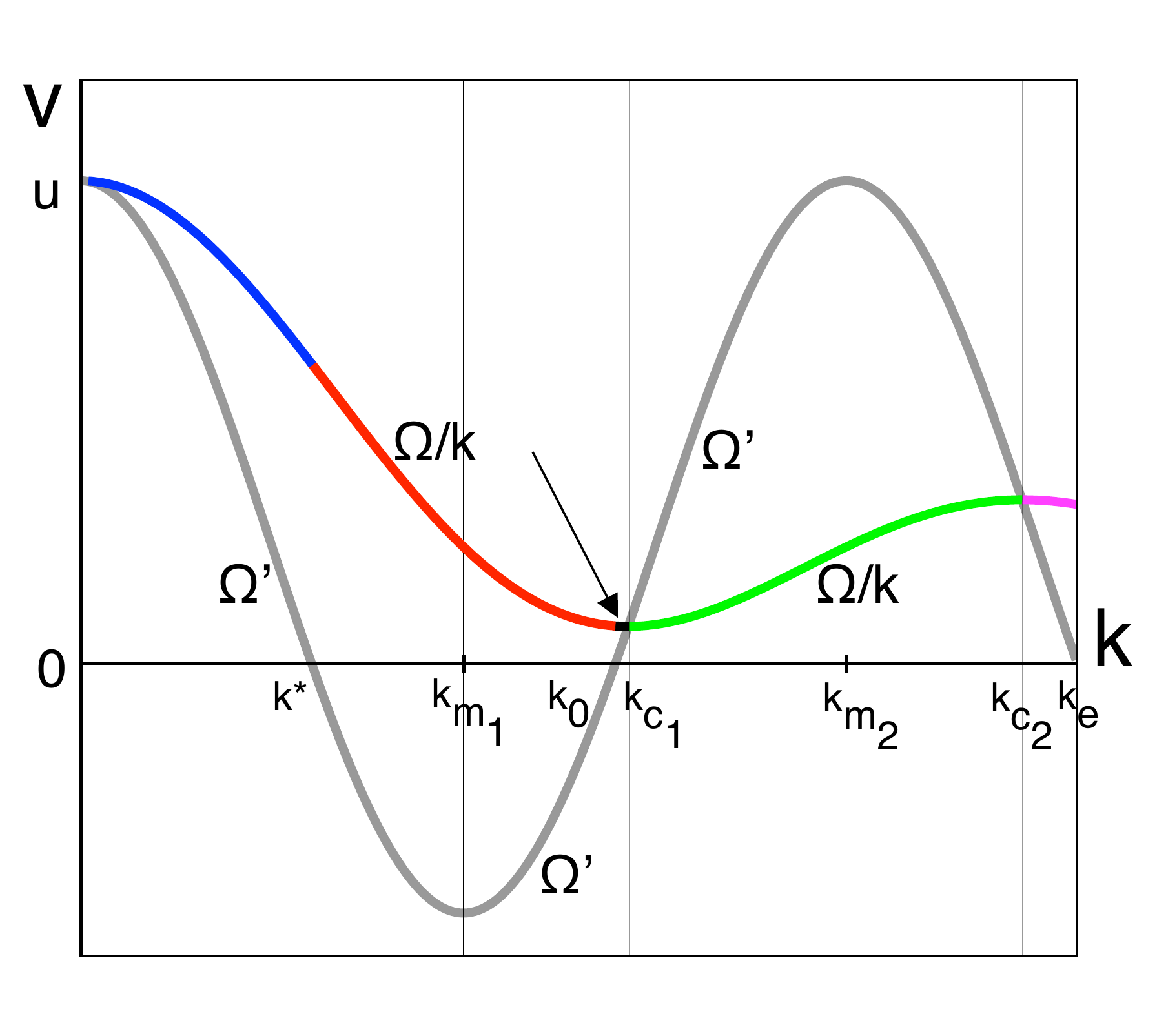} 
\caption{(Color online) Sketches of the group $\Omega'(k)$ and phase $\Omega(k)/k$ velocities of elementary excitations in superfluid $^{4}He$. The extrema of $\Omega(k)$ are located at $k=k^{*}$, $k=k_{0}$ (roton minimum), and $k=k_{e}$ (end point).  Hereafter the function $\Omega=2k + \sin2\pi k$ is employed to mimic the true dispersion law.  Color legend is explained in the main text.} 
\label{hevelocities}
\end{figure}

The dispersion law $\Omega(k)$ of elementary excitations in a superfluid is a non-monotonic function of the wave number $k$ \cite{LL9}:  after an initial linear in $k$ increase (\ref{sound}), the function $\Omega(k)$ reaches a maximum at $k=k^{*}$ followed by a "roton" minimum at $k=k_{0}$.  Therefore the group velocity is negative and takes on its minimal value in the $[k^{*};k_{0}]$ range.  Additionally, the spectrum has an end point $k=k_{e}$ where the group velocity vanishes \cite{LL9}.  As a result, the group velocity is positive and takes on its maximal value in the $[k_{0};k_{e}]$ range.  Dependences of the group and the phase velocities on $k$ are sketched in Figure \ref{hevelocities};  the phase velocity also has a minimum and a maximum at $k$ finite, and both velocities have simultaneous maxima of magnitude $u$ at $k=0$. 
A color legend is employed for the phase velocity curve to distinguish, according to the inequalities (\ref{phases}), five different spectral ranges corresponding to five families of the wavefronts:  

(i)  The $[0;k^{*}]$ range (blue).  Here the group velocity is smaller than the phase velocity.

(ii)  The $[k^{*};k_{0}]$ range (red).  Here the group velocity is negative and takes on a minimal value.

(iii) The $[k_{0};k_{c_{1}}]$ range (black).  Here the group velocity is smaller than the phase velocity.  In the superfluid this range of the wave numbers \cite{GJJ} is very narrow;  an arrow in Figure \ref{hevelocities} points to it.

(iv)  The $[k_{c_{1}};k_{c_{2}}]$ range (green).  Here the group velocity is larger than the phase velocity and the former takes on a maximal value.

(v)  The $[k_{c_{2}}; k_{e}]$ range (magenta).  Here the group velocity is smaller than the phase velocity.

Employing the empirically known dispersion law \cite{LL9}, the extremal phase and group velocities (extrapolated to zero pressure) characterizing the splash in a superfluid can be estimated as:
\begin{eqnarray}
\label{extremal velocities in He}
v_{m_{1}}&=&-2\cdot10^{4}~\text{cm/s},~~~~~~~~v_{c_{1}}=5.9\cdot10^{3}~\text{cm/s},\nonumber\\
v_{m_{2}}&\simeq& u=2.4\cdot10^{4}~\text{cm/s},~~~v_{c_{2}}=9\cdot10^{3}~\text{cm/s}.
\end{eqnarray}
The velocity $v=v_{c_{1}}$ known as the Landau critical roton velocity to destroy superfluidity is also a threshold velocity for the generation of a wake pattern behind a small source uniformly moving through the superfluid \cite{GJJ}.

\begin{figure}
\includegraphics[width=1.0\columnwidth, keepaspectratio]{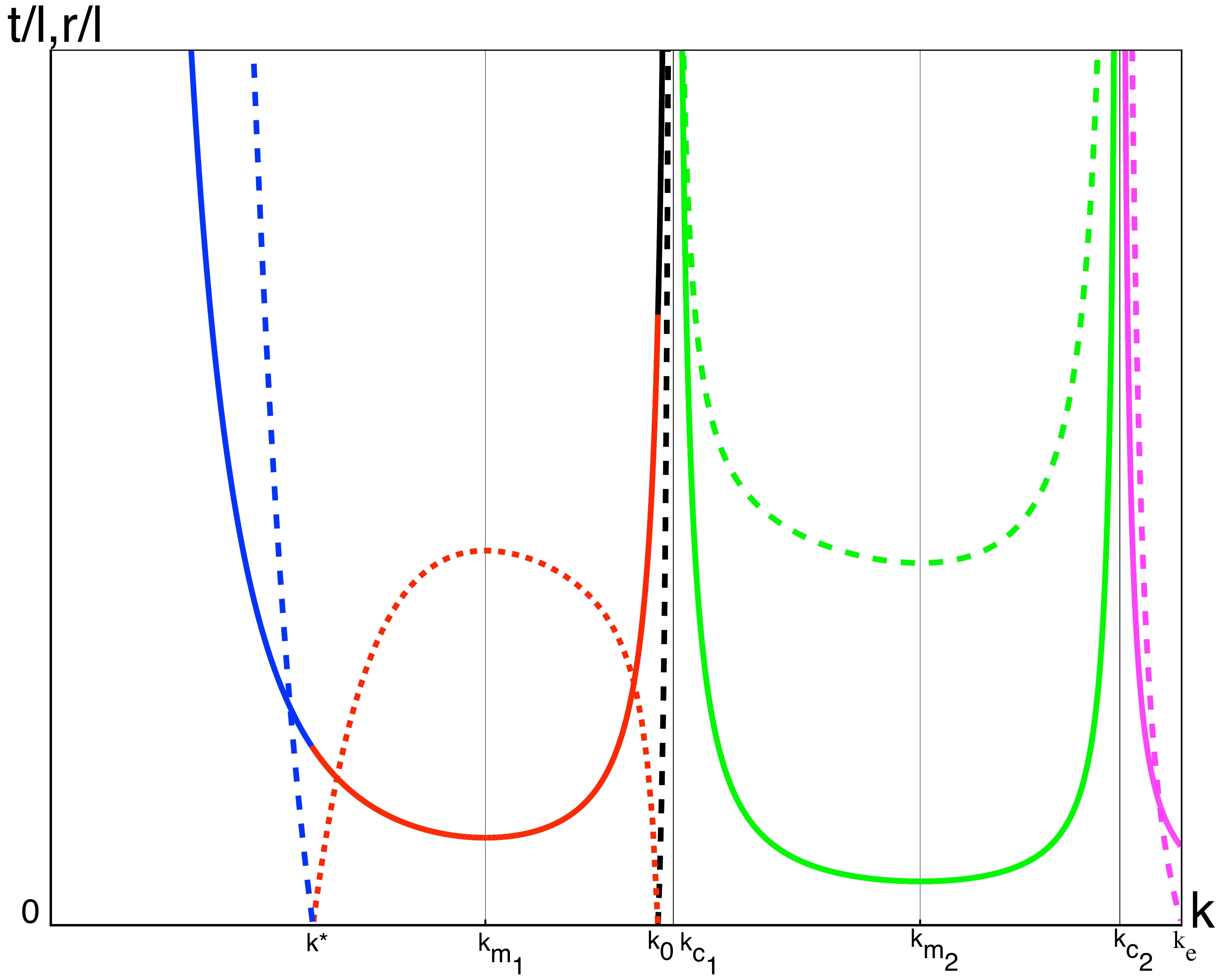} 
\caption{(Color online) The $t(k)/l$ and $r(k)/l$ dependences (\ref{splash_parametric_radial}) and (\ref{phases}) based on the behavior of the velocities in Figure \ref{hevelocities} using the same legend and colors.  Dashed curves of the same color display the $r(k)/l$ dependences.} 
\label{rtofk}
\end{figure}

To understand the dynamics of the wavefronts, in Figure \ref{rtofk} we sketched the $t(k)/l$ and $r(k)/l$ dependences (\ref{splash_parametric_radial}) and (\ref{phases}) color coordinated with Figure \ref{hevelocities}.
The resulting splash pattern in the $(r/l,t/l)$ variables consisting of five families of wavefronts is shown in Figure \ref{hesplash}:
\begin{figure}
\includegraphics[width=1.0\columnwidth, keepaspectratio]{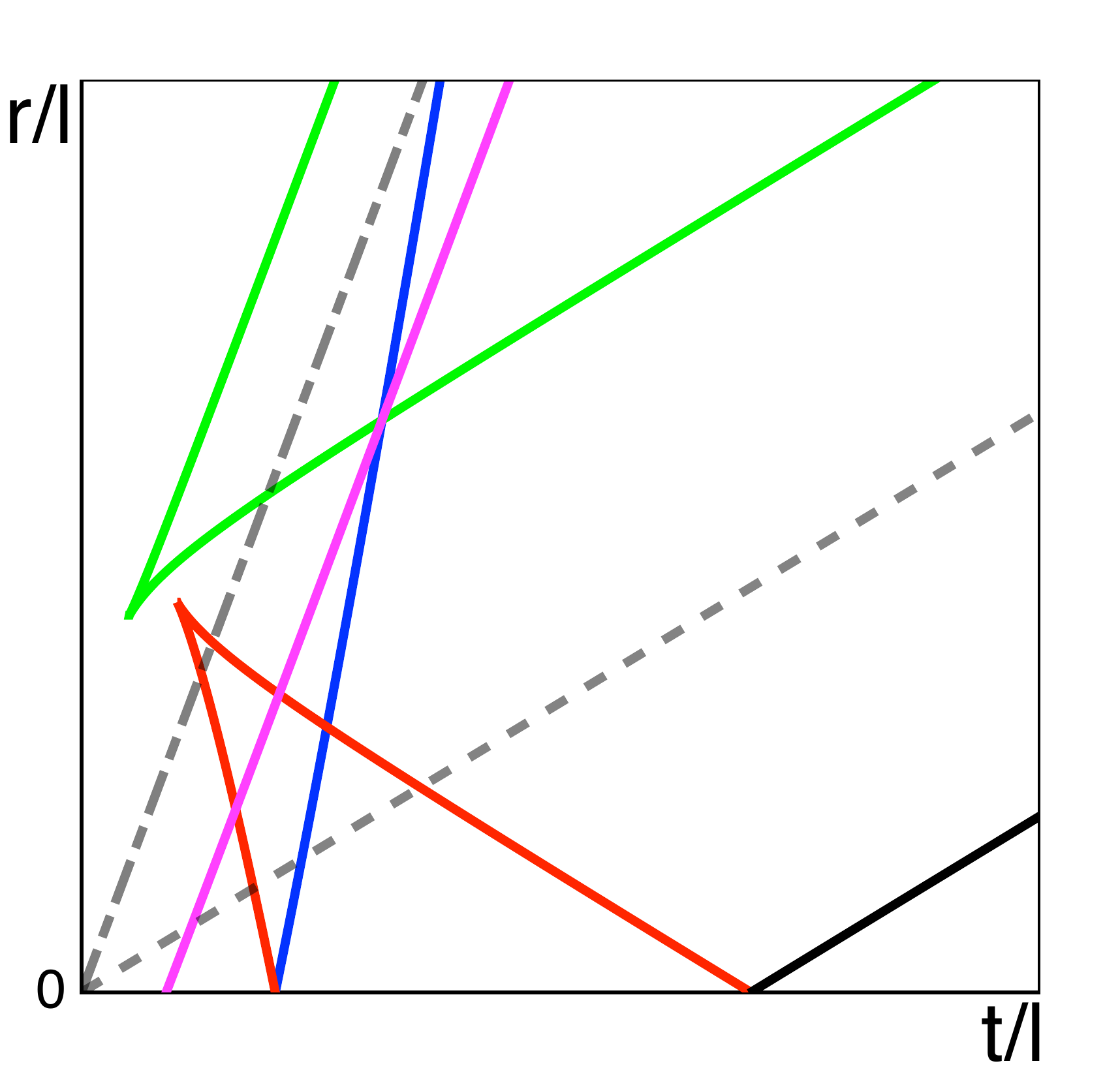} 
\caption{(Color online) Radii of the spherical density wavefronts vs time (scaled by integer factor $l$ to "collapse" wavefronts of given family onto a single curve or a pair of curves) in a superfluid following a sudden localized perturbation, according to Eqs.(\ref{splash_parametric_radial}), (\ref{phases}), and the functional dependences of the phase and group velocities depicted in Figure \ref{hevelocities}.  The color legend is coordinated with Figures \ref{hevelocities} and \ref{rtofk}.  Greyscale dashed and double-dashed lines, $r=v_{c_{1}}t$ and $r=v_{c_{2}}t$, respectively, are shown for reference.} 
\label{hesplash}
\end{figure}

(i) The blue colored (largest slope) wavefronts emerge with period $t_{*}=2\pi/\Omega(k^{*})$ at the center of the splash and expand, asymptotically reaching the speed of sound $u$ (\ref{extremal velocities in He}) for large $t$. 

(ii)  The magenta colored (second largest slope) wavefronts nucleate at the center of the splash with period $t_{e}=2\pi/\Omega(k_{e})$ (\ref{interval}).  As $t\rightarrow \infty$, the corresponding solution to the equation for $t(k)$ (Figure \ref{rtofk}) approaches $k=k_{c_{2}}$, thus implying that the wavefronts expand asymptotically reaching the velocity $v=v_{c_{2}}$ (\ref{extremal velocities in He});  the greyscale double-dashed line $r=v_{c_{2}}t$ is shown for reference.  

(iii) The black colored (smallest slope) wavefronts emerge at the center of the splash with period $t_{0}=2\pi/\Omega(k_{0})$ and expand, asymptotically reaching the Landau critical roton velocity $v=v_{c_{1}}$ (\ref{extremal velocities in He}) for large $t$; the  greyscale dashed line $r=v_{c_{1}}t$ is shown for reference. 

(iv) The dynamics of the green colored (diverging) wavefronts can be understood via the argument already given in the discussion of the water splash.  They  nucleate at $r$ finite with period $t_{m_{2}}$ (\ref{interval}); the locus of these events belongs to the straight line $r=v_{m_{2}}t$.  Each nucleation event results in a pair of diverging spherical wavefronts.  In the large time limit the slower of the two expands with velocity approaching the Landau critical roton velocity $v=v_{c_{1}}$ (\ref{extremal velocities in He}), while the faster expands with velocity approaching $v=v_{c_{2}}$ (\ref{extremal velocities in He}).

(v)  The red colored (\textit{converging}) wavefronts made by elementary waves of negative group velocity nucleate at $r$ finite with period $t_{m_{1}}$ (\ref{interval}); the locus of these events belongs to the straight line $r=|v_{m_{1}}|t$.  Since now $r(k)/l$ has a maximum at $k=k_{m_{1}}$, each nucleation event results in a pair of converging spherical wavefronts.  The faster of these reaches the center of the splash exactly as the blue colored (largest slope) wavefronts nucleate there while the slower one arrives at the center when black colored (smallest slope) wavefronts emerge there.  This lends itself to the following interpretation:  converging wavefronts made of waves of negative group velocity rebound off the splash center in the form of diverging wavefronts made of waves of positive group velocity.  

\section{Conclusions}    

To summarize, we demonstrated that in large time regime dispersion law of relevant collective excitations alone suffices to understand dynamics of wavefronts in splashes in isotropic media.   The outcome is determined by the interplay between excitation's phase and group velocities as well as the sign of the latter. The salient features of splashes are controlled by the existence of extremal values of the phase and the group velocities: the group velocity gives the expansion rate of the locus of the points where new wavefronts nucleate or existing ones disappear, while the phase velocity determines the large-time expansion rate of a group of wavefronts.  If the group velocity is negative in a spectral range and takes on a minimal value within it, then \textit{converging} wavefronts will be present in the splash.   

To illustrate our theory we also carried out several case studies of experimentally relevant setups.  Specifically, splashes on water and in two-dimensional electron gas were found to be similar: following a localized perturbation, a quiescent region inside the annular waves forms.  This region expands with a constant rate corresponding to the minimum group velocity -- the speed of sound for the two-dimensional electron gas or $18~cm/s$ for water, the conclusion due to Rayleigh \cite{Rayleigh}.  New wavefronts nucleate at the boundary of the quiescent region at regular time intervals ($0.43~s$ for water), in pairs (in water) or one at a time (in a two-dimensional electron gas).  When the wavefronts appear in pairs, one of them expands with the minimal phase velocity ($23~cm/s$ for water).  The other (water and two-dimensional electron gas) expands asymptotically with a constant acceleration.  The gross features of a splash in a superfluid are determined by \textit{five} extremal velocities.  Additionally, due to the existence of a negative group velocity spectral range, some of the wavefronts in superfluid $^{4}He$ splash are converging.    

The existence of converging wavefronts in splashes is not unique to superfluid $^{4}He$.  Similar conclusions apply to a dipolar quantum gas whose spectrum also features a roton minimum \cite{dipole}.  More generally, whenever there exists a spectral range where the group velocity as a function of the wavenumber is negative and it takes on a minimal value within this range, then converging wavefronts will be necessarily present in the splash.  The first known realistic example of a spectrum featuring excitations of negative group velocity, the optical branch of vibrations in crystals \cite{history}, belongs to this category, too.

We hope that both the general analysis and sample studies carried out in this work will guide future observations of splashes.  

\section{Acknowledgements}

The author is grateful to J. P. Straley and A. P. Levanyuk for valuable comments.


\begin{thebibliography}{33}

\bibitem{He41}  T. Zhang and S.W. Van Sciver, \textit{Large-scale turbulent flow around a cylinder in counterflow superfluid He (He(ii))}, Nat. Phys. \textbf{1}, 36 (2005).

\bibitem{He42}  G. P. Bewley, D. P. Lathrop, and K. R. Sreenivasan, \textit{Visualization of quantized vortices}, Nature \textbf{441}, 588 (2006).

\bibitem{He43}  W. Guo, S. B. Cahn, J. A. Nikkel, W. F. Vinen, and D. N. McKinsey, \textit{Visualization study of counterflow in superfluid $^{4}He$ using metastable helium molecules}, Phys. Rev. Lett. \textbf{105}, 045301 (2010).

\bibitem{He44}  W. Guo, M. La Mantia, D. P. Lathrop and S. W. Van Sciver, \textit{Visualization of two-fluid flows of superfluid helium-4}, Proc. Natl Acad. Sci. USA \textbf{111}, 4653 (2014).

\bibitem{He3}  S. N. Fisher, M. J. Jackson, Y. A. Sergeev, and V. Tsepelin, \textit{Andreev reflection, a tool to investigate vortex dynamics and quantum turbulence in $^{3}He-B$}, Proc. Natl Acad. Sci. USA \textbf{111}, 4659 (2014).

\bibitem{vapor}  A. Kumar, N. Anderson, W. D. Phillips, S. Eckel, G. K. Campbell, and S. Stringari, \textit{Minimally destructive, Doppler measurement of a quantized flow in a ring-shaped Bose-Einstein condensate}, N. J. Phys. \textbf{19}, 025001 (2016).

\bibitem{imaging} J.-P. Tetienne, N. Dontschuk, D. A. Broadway, A. Stacey, D. A. Simpson, L. C. L. Hollenberg, \textit{Quantum imaging of current flow in graphene}, Sci. Adv. \textbf{3}, e1602429 (2017).

\bibitem{super} X. Liu,  Y. X. Chong, R. Sharma, and J. C. S\'eamus Davis, \textit{Atomic-scale visualization of electronic fluid flow}, Nature  Materials, \textbf{20}, 1480 (2021).

\bibitem{Levitov}  L. Levitov and G. Falkovich, \textit{Electron viscosity, current vortices and negative
nonlocal resistance in graphene}, Nat. Phys. \textbf{12}, 672 (2016).

\bibitem{Kelvin_Mach}  E. B. Kolomeisky and J. P. Straley, \textit{Kelvin-Mach Wake in a Two-Dimensional Fermi Sea}, Phys. Rev. Lett. \textbf{120}, 226801 (2018).

\bibitem{Lamb} H. Lamb, \textit{Hydrodynamics} (6th ed., Cambridge University Press, 1975), Chapter IX.

\bibitem{Crapper}  G. D. Crapper, \textit{Introduction to water waves} (Ellis Horwood Limited, 1984), Chapter 5, specifically Figures 5.2 and 5.5.

\bibitem{Stoker}  J. J. Stoker, \textit{Water Waves: The Mathematical Theory with Applications} (Dover Publications, 2019), Chapter 6, specifically, Figure 6.6.2,

\bibitem{Ferrell}  R.A. Ferrell, \textit{Long Lifetime of Positronium in Liquid Helium}, Phys. Rev. \textbf{108}, 167
(1957).

\bibitem{Eva_Andrei}  J. Mao, Y. Jiang, D. Moldovan, G. Li, K. Watanabe, T. Taniguchi, M. R. Masir, F. M. Peeters and E. Y. Andrei, Nature Physics, \textit{Realization of a tunable artificial atom at a supercritically charged vacancy in graphene}, \textbf{12}, 545 (2016).

\bibitem{LL6} L. D. Landau and E. M. Lifshitz, \textit{Fluid Mechanics} (Pergamon, Oxford, 1987), Sections 12, 61 and 62.

\bibitem{Kelvin}  W. Thomson, \textit{On the Waves Produced by a Single Impulse in Water of Any Depth, or in a Dispersive Medium}, Proc. R. Soc. Lond.\textbf{42}, 80-83 (1887).

\bibitem{Rayleigh}  L. Rayleigh, \textit{Hydrodynamical notes}, Philos. Mag. \textbf{21}, 177 (1911).

\bibitem{Lighthill}   J. Lighthill, \textit{Waves in Fluids} (Cambridge University Press, Cambridge, UK, 1978), Chapter 3.

\bibitem{Le}  B. Le M\'ehaut\`e, \textit{Gravity-capillary rings generated by water drops}, J. Fluid Mech. \textbf{197}, 415 (1988).

\bibitem{Fetter}  A. L. Fetter,  \textit{Electrodynamics of a Layered Electron Gas. I. Single Layer},  Annals of Physics \textbf{81}, 367 (1973).

\bibitem{AFS}  T. Ando, A. B. Fowler and F. Stern, \textit{Electronic properties of two-dimensional systems}, Rev. Mod. Phys. \textbf{54}, 437 (1982), and references therein.

\bibitem{Hwang_D_Sarma}  E. H. Hwang and S. Das Sarma, \textit{Dielectric function, screening, and plasmons in two-dimensional graphene}, Phys. Rev. B \textbf{75}, 205418 (2007). 

\bibitem{DasHwang}  S. Das Sarma and E. H. Hwang, \textit{Collective Modes of the Massless Dirac Plasma}, Phys. Rev. Lett. \textbf{102}, 206412 (2009).

\bibitem{EBKJPS}  E. B. Kolomeisky and J. P. Straley, \textit{Screening and plasma oscillations in an electron gas in the hydrodynamic approximation}, Phys. Rev. B \textbf{96}, 165116 (2017).

\bibitem{LL9}  E. M. Lifshitz and L. P. Pitaevskii,  \textit{Statistical Physics}, Third edition, Part 2: Volume 9 (Course of Theoretical Physics) (Butterworth-Heinemann, 1980), Chapters III and IX.

\bibitem{GJJ}  E. B. Kolomeisky, J. Colen and J. P. Straley, \textit{Negative group velocity and Kelvin-like wake pattern}, Phys. Rev. B \textbf{105}, 054509 (2022).

\bibitem{LL5}  L. D. Landau and E. M. Lifshitz, \textit{Statistical Physics}, Third edition, Part 1: Volume 5 (Course of Theoretical
Physics) (Butterworth-Heinemann, 1980), Section 123.

\bibitem{Pines_Nozieres}  D.Pines and P. Nozi\`eres, \textit{The Theory of Quantum Liquids}, Vol.1 (Benjamin, New York 1966), Chapter 2.

\bibitem{JCEBK} J. Colen and E. B. Kolomeisky, \textit{Kelvin-Froude wake patterns of a traveling pressure disturbance}, Eur. J. Mech. /B Fluids \textbf{85}, 400 (2021). 

\bibitem{Atkins}  K. R. Atkins and Y. Narahara, \textit{Surface Tension of Liquid $He^{4}$}, Phys. Rev. \textbf{138}, A437 (1965).

\bibitem{dipole}  L. Chomaz, R. M. W. van Bijnen, D. Petter, G. Faraoni, S. Baier, J. H. Becher, M. J. Mark, F. W\"achtler, L. Santos and F. Ferlaino, \textit{Observation of roton mode population in a dipolar quantum gas}, Nature Physics \textbf{14}, 442 (2018), and references therein.

\bibitem{history} A brief history of negative group velocity is outlined in K. T. McDonald, \textit{Negative group velocity}, Am. J. Phys. \textbf{69}, 607 (2001).

\end{thebibliography}
\end{document}